\documentclass[usenatbib,usegraphicx]{mn2e}

\usepackage{graphicx}
\usepackage{amsmath}
\usepackage{url}
\usepackage{color}

\title{Observational biases in flux magnification measurements}

\author[H.~Hildebrandt et al.]{H.~Hildebrandt,$^{1}$\thanks{Email: hendrik@astro.uni-bonn.de}
\\
$^1$Argelander-Institut f\"ur Astronomie, Auf dem H\"ugel 71, 53121 Bonn, Germany\\
}

\date{Released 2014}

\pagerange{\pageref{firstpage}--\pageref{lastpage}} \pubyear{2014}

\begin{document}
\setlength{\voffset}{-12mm}

\label{firstpage}

\maketitle
\begin{abstract}
  Flux magnification is an interesting complement to shear-based
  lensing measurements, especially at high redshift where sources are
  harder to resolve. One measures either changes in the source density
  (magnification bias) or in the shape of the flux distribution
  (e.g. magnitude-shift). The interpretation of these measurements
  relies on theoretical estimates of how the observables change under
  magnification. Here we present simulations to create multi-band
  photometric mock catalogues of Lyman-break galaxies in a
  CFHTLenS-like survey that include several observational effects that
  can change these relations, making simple theoretical estimates
  unusable. In particular, we show how the magnification bias can be
  affected by photometric noise, colour selection, and dust
  extinction. We find that a simple measurement of the slope of the
  number-counts is not sufficient for the precise interpretation of
  virtually all observations of magnification bias. We also explore
  how sensitive the shift in the mean magnitude of a source sample in
  different photometric bands is to magnification including the same
  observational effects. Again we find significant deviations from
  simple analytical estimates. We also discover a
  wavelength-dependence of the magnitude-shift effect when applied to
  a colour-selected noisy source sample. Such an effect can mimic the
  reddening by dust in the lens. It has to be disentangled from the
  dust extinction before the magnitude-shift/colour-excess can be used
  to measure the distribution of either dark matter or extragalactic
  dust. Using simulations like the ones presented here these
  observational effects can be studied and eventually removed from
  observations making precise measurements of flux magnification
  possible.
\end{abstract}

\section{Introduction}
\label{sec:intro}
The magnification effect of weak gravitational lensing is being used
in an increasing number of projects to study dark matter structures,
ranging from galaxy halos
\citep{2005ApJ...633..589S,2009A&A...507..683H,2010MNRAS.405.1025M,2014MNRAS.440.3701B}
over groups and clusters of galaxies
\citep{2010ApJ...723L..13V,2011ApJ...733L..30H,2012ApJ...754..143F,2014MNRAS.439.3755F,2011ApJ...729..127U,2014ApJ...795..163U}
to the large-scale-structure of the Universe
\citep{2010MNRAS.401.2093V,2011MNRAS.415.1681H,2012MNRAS.426.2489M}. In
contrast to the widely used shear effect of weak lensing some
magnification observables do not require the spatial resolution of the
source galaxies used in the measurement. These flux magnification
observables offer the advantage of a larger source density as well as
a wider redshift range (especially higher redshifts) that can be
accessed.

Theoretical modelling of these flux magnification observables relies
on the knowledge of the flux distribution, hence the magnitude
number-counts, of the sources, before magnification. The commonly used
\emph{magnification bias} and \emph{magnitude shift} effects represent
changes in the first two moments of this flux distribution. Most
studies so far have assumed knowledge of the intrinsic flux
distribution or that the observed flux distribution averaged over wide
areas on the sky is an unbiased estimate of the intrinsic
distribution. The observables can then be calculated from lensing
theory since magnification changes this intrinsic distribution in a
characteristic way.

The actual observed flux distribution, however, can be affected by a
number of different systematic effects. In this paper we show that
depending on the observational technique these effects can be large
and render the use of the observed distribution in combination with
simple analytical models inaccurate. In this paper we explore such
effects with simple simulations of multi-band photometric mock
catalogues and quantify their importance in realistic scenarios. We
also describe a way forward to interpret flux magnification
measurements in the future.

The paper is organised as follows. We present the theoretical
background and analytical modelling of the flux magnification
quantities in Sect.~\ref{sec:theory}. The systematic effects that we
simulate are discussed in Sect.~\ref{sec:systematics}, and the
simulations are described in Sect.~\ref{sec:simulations}. Results are
presented in Sect.~\ref{sec:results} and discussed in
Sect.~\ref{sec:discussion} before we summarise and conclude in
Sect.~\ref{sec:summary}.

\section{Theoretical background}
\label{sec:theory}
\subsection{Magnification bias}
Source samples of objects used in weak gravitational lensing studies
(typically galaxies but also QSOs) are always selected on one or more
observables. This is necessary to avoid regions of parameter space
where the measurements are unreliable (e.g. faint magnitudes, small
sizes). Magnification by gravitational lensing can change these
observables, e.g. flux and size. The underlying reason for this is
that lensing stretches solid angles on the sky. In combination these
effects can lead to a net increase or decrease in the number of
objects in a sample. This effect is called magnification bias.

The best-known example of magnification bias is the change in the
number density of a magnitude-limited sample of QSOs
\citep{2005ApJ...633..589S} or galaxies
\citep{2009A&A...507..683H}. Magnification is pushing objects over a
fixed magnitude limit. If this increase in object density is larger
than the decrease due to the stretching of the sky one observes an
over-density of objects in magnified regions. Otherwise one might see a
decrease. Only if the intrinsic distribution (i.e. the distribution
before lensing) of fluxes is shaped in such a way that it exactly
compensates both effects, one does not observe a change in density.

The density change depends on the logarithmic slope $\alpha_{\rm nc}$
of the differential magnitude number-counts, $n(m)$,\footnote{Those
  should not be confused with the cumulative number-counts $N(m)$
  often used in the literature.} of the source sample \citep[see
also][]{2009A&A...507..683H}:
\begin{equation}
  \label{eq:alpha_nc}
  \alpha_{\rm nc}(m)\equiv2.5\frac{{\rm d}\log\left[n(m)\right]}{{\rm d}m}\,.
\end{equation}
For a narrow source redshift slice the shape of the number-counts is
equivalent to the shape of the luminosity function.

Under magnification the number-counts are changed in the following
characteristic way:
\begin{equation}
  \label{eq:nc_mu}
  n(m)=\mu^{-1}\,n_0\left[m+2.5\log(\mu)\right]\,,
\end{equation}
with $\mu$ being the magnification and $n_0$ being the intrinsic
number-counts before lensing. Through Taylor expansion it can easily
be shown that
\begin{equation}
  \label{eq:nc_mu_alpha}
  n(m)=\mu^{\alpha_{\rm nc}-1}n_0(m)\approx[1+(\alpha_{\rm nc}-1)\,\delta\mu]\, n_0(m)\,,
\end{equation}
where the approximation on the right-hand side holds in the weak
lensing regime, i.e. $|\delta\mu|\equiv|\mu-1|\ll1$.

Using the relations in either Eq.~(\ref{eq:nc_mu}) or
Eq.~(\ref{eq:nc_mu_alpha}) requires perfect knowledge of the
number-counts. In practice the measurement of the number-counts is
always affected by noise. It is usually also assumed that properties
of the source sample such as their colour distribution do not change
under magnification. As we will see in the remainder of the paper
these assumptions are not always justified.

In practice one needs to consider a source population within a finite
magnitude interval $[m_{\rm a}; m_{\rm b}]$. The magnification bias then becomes a
measurement of the change of the zeroth order moment of the
number-counts in this interval:
\begin{equation}
  I_0=\int_{m_{\rm a}}^{m_{\rm b}}\,n(m)\,{\rm d}m\,.
\end{equation}
According to Eq.~(\ref{eq:nc_mu_alpha}) this integrated quantity changes
under magnification if $\alpha_{\rm nc} \neq 1$.\footnote{This assumes
that $\alpha_{\rm nc}$ is constant over the interval $[m_{\rm a};
m_{\rm b}]$.}

\subsection{Magnitude shift}
Instead of integrating over the number-counts as in the case of the
magnification bias one can also look at changes in the higher-order
moments. A measurement of the shift in the mean magnitude corresponds
to detecting a change of the first moment divided by the zeroth
moment:
\begin{equation}
  \left<m\right>=\frac{I_1}{I_0}=\frac{\int_{m_{\rm a}}^{m_{\rm b}}m\:n(m)\,{\rm d}m}{\int_{m_{\rm a}}^{m_{\rm b}}\,n(m)\,{\rm d}m}\,.
\end{equation}
Such a change can only be detected if $\frac{{\rm d}^2\log n(m)}{{\rm
    d}m^2}\neq0$, i.e. if the logarithmic number-counts show some
curvature over the magnitude interval under consideration. Hence, for
number-counts that follow a power law with a constant slope the mean
magnitude does not change under magnification.

As \cite{2010MNRAS.405.1025M} showed the magnitude shift can be
linearised and approximated in the weak lensing regime by:
\begin{equation}
\label{eq:mag_shift_linear}
  \delta m_{\rm obs}=\left<m\right>-\left<m_0\right>=C_{\rm S}\, \delta m_{\rm ind}\,,
\end{equation}
where $\delta m_{\rm obs}$ is the observed magnitude shift,
\mbox{$\delta m_{\rm ind}=-2.5\log{\mu}$} is the induced magnitude
shift, and the constant $C_{\rm S}$ can be calculated:
\begin{equation}
  \label{eq:C_S}
  C_{\rm S}= 1 - \frac{1}{N_{\rm tot}} \Big\{ \big[n_{\rm b}m_{\rm b}-n_{\rm a}m_{\rm a}\big] - \left<m_0\right> \big[n_{\rm b}-n_{\rm a}\big] \Big\}\,,
\end{equation}
with $N_{\rm tot}=I_0$ and $n_{a,b}\equiv n(m_{a,b})$.\footnote{Note
  that \cite{2010MNRAS.405.1025M} use just a faint magnitude cut
  whereas we use a bright and a faint cut here.} The constant $C_S$
expresses the factor by which the observed magnitude shift of an
ensemble of objects is suppressed to the (unobservable) magnitude
shift of an individual object. This factor depends on the curvature of
the logarithmic number-counts. It vanishes for number counts that
follow a power law.

\subsection{Reddening}
Since gravitational lensing is fundamentally achromatic one would not
expect to see a change in colour from lensing alone. It was suggested
by \cite{2010MNRAS.405.1025M} to use the difference in the magnitude
shifts in different bands as an estimate of reddening due to dust that
is associated with the gravitational lens.

This approach is only valid if the galaxies magnified into the sample
and magnified out of the sample have the same colour distribution. In
the presence of noise, non-detections in some bands, and colour
pre-selection this is not always the case as we will show in the
following.

If one is only interested in lensing then dust reddening represent a
systematic effect that needs to be removed. However, in this paper we
do not treat it as a systematic effect for two reasons. First, the
other systematic effects discussed here are all observational in
nature whereas the dust is an astrophysical effect that alters flux
magnification observables. Secondly, one can also regard the dust
reddening as the main observable exploiting the fact that the
techniques described here represent unique ways to study dust on large
scales. For these reasons we treat dust as an observable and not as a
systematic in this paper.

\section{Observational effects}
\label{sec:systematics}
\subsection{Incompleteness}
The sources used in magnification measurements are extracted from
noisy data. Whether a celestial source enters a catalogue or not
depends on several factors. In optical astronomy source detection
means typically a combination of a flux threshold and a
surface-brightness threshold \citep{1996A&AS..117..393B}. At fixed
flux two objects can have widely different sizes and hence different
surface-brightness. Thus, when operating close to the noise limit (and
one almost always has to do this in weak lensing applications to reach
a sufficient source density) one will preferentially miss those
objects with the lowest surface-brightness at a given flux. This is
one reason for incompleteness. 

In this paper we are simulating a high-redshift source sample whose
galaxies are usually unresolved in ground-based data. This is a case
of particular interest for magnification since it allows us to measure
gravitational lensing in a regime that is not accessible by
shear-based methods that rely on ellipticity measurements of
well-resolved source galaxies. Since we assume that our sources are
unresolved we ignore surface-brightness effects in the following. It
should be noted though that this can become important for
flux-magnification measurements employing lower-$z$ source samples.

Another reason for incompleteness is photon shot noise randomly
scattering objects below or above the flux/surface-brightness
threshold. For a non-flat flux distribution this leads to Eddington
bias \citep{1913MNRAS..73..359E,2004A&A...424...73T}. Due to their
larger abundance more faint objects are scattered to brighter
magnitudes than vice versa. The simulations used here account for this
effect by construction.

\subsection{Colour selection}
A major problem when measuring magnification bias is the separation of
sources and lenses in redshift. If the source and lens samples overlap
in redshift this will give rise to physical cross-correlations in
their angular positions. These physical cross-correlations are
typically much larger than the magnification bias signal one is
interested in and hence one needs to carefully separate the samples in
redshift to minimise this effect. Redshift separation can be achieved
with different techniques, most of them involving colour cuts. Note
that separating galaxies by photo-$z$ is not conceptually different
than cutting in colour space. The photo-$z$ method represents just a
more complicated, higher-dimensional, and usually less-transparent
colour-cut.

Redshift separation by colour selection makes use of the fact that
different regions in colour space map to different redshift
regions. Applying hard cuts in colour, however, will yield a source
sample that is intrinsically different (e.g. in terms of its redshift
distribution) at different fluxes due to the different noise
levels. Colour space is not evenly distributed and the density of
galaxies can show strong gradients. Introducing noise will
asymmetrically scatter galaxies around colour space generally
decreasing these gradients. Hence, a fixed colour cut will typically
yield a different source sample for faint galaxies than for bright
ones with the former having a wider redshift distribution (and
potentially a larger number of outliers at unwanted redshifts) than
the latter. Conversely, galaxies at fixed redshift will be spread out
over a larger region in colour space when there is more noise so that
a fixed colour cut targeting these galaxies will lose an increasing
fraction of the sample for decreasing signal-to-noise ratio (S/N).

In this study we simulate source galaxies at a fixed redshift. So we
ignore the effect of galaxies at other redshifts scattering into the
colour selection box \cite[i.e. the region of colour space targeted to
select the source sample; see Fig.~2 of][for an example of such a
box]{2009A&A...498..725H}. However, we account for the effect of noise
scattering objects out of the selection box. In this particular case -
faint high-$z$ galaxies as background sources - this is the more
serious effect since our sources are noisy compared to the main sample
of galaxies and the $u$-dropouts simulated here (see
Sect.~\ref{sec:simulations}) are generally a very clean high-redshift
sample. In more general applications both effects can be of equal
importance and need to be taken into account.

\section{Simulation setup}
\label{sec:simulations}
For the purpose of this work we assume a background source sample
consisting of Lyman-break galaxies
\citep[LBGs,][]{1996ApJ...462L..17S, 2003ApJ...592..728S,
  2002ARA&A..40..579G} as it was used in the flux magnification
studies \cite{2009A&A...507..683H, 2011ApJ...733L..30H,
  2013MNRAS.429.3230H}, \cite{2012MNRAS.426.2489M}, and
\cite{2012ApJ...754..143F, 2014MNRAS.439.3755F}. Some of the selection
effects studied here are particularly strong when such a
colour-selected, noisy sample is used. It should be noted, however,
that any faint, photo-$z$ selected sample of background sources
\citep[like e.g. the QSOs used in][]{2005ApJ...633..589S,
  2010MNRAS.405.1025M} will show similar effects since the photo-$z$
selection is essentially a more complicated cut in multi-dimensional
colour space. The LBGs used here have the advantage that the colour
selection is fairly simple and transparent whereas the inner workings
of a photo-$z$ code lead to a more complicated selection that is
harder and more time-consuming to simulate.

Some of the effects studied here depend on the level of noise in the
data. Here we simulate a survey similar in its noise properties to the
CFHTLenS data set. This five-band imaging survey carried out with
MegaCam@CFHT has been extensively described in
\cite{2012MNRAS.427..146H}, \cite{2012MNRAS.421.2355H},
\cite{2013MNRAS.433.2545E}, and \cite{2013MNRAS.429.2858M}. We take
the limiting magnitudes in the five optical bands $ugriz$ from Table~1
of \cite{2013MNRAS.433.2545E}.

For simplicity we assume here that we have a pure, uncontaminated
source sample, i.e. all sources are at the same redshift behind the
lenses. We choose the redshift of the sources to be $z=3.2$ identical
to the mean redshift of the $u$-dropouts studied in
\cite{2009A&A...498..725H,2009A&A...507..683H}. The colour selection
criteria for this sample are:
\begin{align}
\label{eq:colour_selection}
1.5 &<(u-g) \nonumber\\ -1.0 &< (g-r)<1.2 \\ 1.5\cdot(g-r) &< (u-g)-0.75\nonumber
\end{align}

From the CFHTLS-Deep field data \citep{2009A&A...498..725H} we also
estimate the mean colours of the LBGs in the magnitude range of
interest. These data are deeper than our simulated data set ($\sim
2.5$mag) such that we do not expect any inaccuracies in the colour
estimates due to possible non-detections, even in the $u$-band, over
the magnitude range of interest. We also model the colour distribution
as a multi-variate Gaussian with the width in the different dimensions
estimated from the same data.

We further assume that the source sample follows a Schechter
luminosity function:
\begin{equation}
  \Phi(M)=\Phi_0\left[10^{0.4(M_*-M)}\right]^{\alpha_{\rm LF}+1}\cdot\,\exp{\left[-10^{0.4(M_*-M)}\right]} \,,
\end{equation}
with $\Phi_0$ being the overall normalisation, $M_*$ being the
characteristic absolute magnitude, and $\alpha_{\rm LF}$ being the
faint-end slope. We use the parameters for the $u$-dropouts reported
in \cite{2010A&A...523A..74V}. The derivative of the Schechter
function that is under idealised conditions directly related to the
slope $\alpha_{\rm nc}$ (see Eq.~\ref{eq:alpha_nc}) is:
\begin{equation}
\label{eq:Schechter_prime}
  \frac{{\rm d}\Phi}{{\rm d}M}=-0.4\,\ln{(10)}\,\left(\alpha_{\rm LF}+1-10^{0.4(M_*-M)}\right)\,\Phi(M)\,.
\end{equation}

Besides introducing magnification we also consider the existence of
dust which can be present in lens galaxies. This can counter-act the
magnification by dimming background objects and hence affects the
interpretation of a magnification measurement. A Milky-Way dust
extinction law by \cite{1989ApJ...345..245C} is assumed and we place
the lens galaxy at a redshift of $z=0.7$. In combination with the
effective wavelength of the MegaCam@CFHT filter set this allows us to
calculate the change in colour of background objects behind lenses
that contain such dust.

The simulation consists of the following steps:
\begin{itemize}
\item The $r$-band magnitude interval of interest ($22<r<26.5$ in this
  case) is split into steps of $\Delta r=0.001$.
\item Each $r$-band magnitude bin contains 200 objects such that the
  each simulation contains $9\times10^5$ galaxies. For a limited set
  of parameters we also run simulations with 100 times more objects to
  check for possible numerical inaccuracies.
\item Each bin is assigned a weight corresponding to the value of the
  Schechter function at this apparent $r$-band magnitude (converted to
  absolute magnitudes assuming a standard $\Lambda$CDM universe and a
  source redshift of $z=3.2$). This ensures that all magnitudes are
  sampled equally well. If one just randomly picked from a realistic
  Schechter function one would inevitably simulate very few bright
  objects due to the exponential shape of that function whose value
  changes by several orders of magnitude over the magnitude range
  considered here. In the following, all statistics will include those
  weights without mentioning them explicitly again.
\item We consider four scenarios:
  \begin{enumerate}
  \item Reference scenario without magnification and extinction:
    \begin{equation}
      \Phi_{\rm ref}(M)=\Phi(M)\,.
    \end{equation}
  \item Scenario with magnification $\mu$:
    \begin{equation}
      \Phi_\mu(M)=\frac{1}{\mu}\Phi(M+2.5\log_{10}{\mu})\,.
    \end{equation}
  \item Scenario with extinction due to dust with the absorption in
    the $r$-band being $A_r$:
    \begin{equation}
      \Phi_\tau(M)=\Phi(M-A_r)\,.
    \end{equation}
  \item Scenario with magnification and extinction combined:
    \begin{equation}
      \Phi_{\mu+\tau}(M)=\frac{1}{\mu}\Phi(M+2.5\log_{10}{\mu}-A_r)\,.
    \end{equation}
  \end{enumerate}
  Several different values for the magnification $\mu$ as well as the
  dust absorption $A_r$ are simulated. In scenario (iv) we scale the
  value of the absorption in the rest-frame visual, $A_V$, with the
  magnification excess $\delta\mu$:
  \begin{equation}
    \label{eq:A_V}
    A_V=c_{\rm d}\,\delta\mu\,. 
  \end{equation}
  Subsequently we calculate the absorption in the observed-frame
  $r$-band, $A_r$, by assuming the Milky-Way extinction law mentioned
  above. Note that the assumption of the linear relation between $A_V$
  and $\delta\mu$ does not limit the generality of the arguments
  presented in this paper in any way. It is just a convenient choice
  to present the results and relate our findings more directly to
  other work.
\item Magnitudes in the other bands are assigned to each $r$-band
  magnitude bin according to the mean colours of the LBGs in the
  reference survey. For scenarios that include extinction the colours
  are modified according to the extinction law described above.
\item For scenarios that include an intrinsic distribution of colours
  of the LBGs, random Gaussian colour offsets are calculated for each
  object and added to the magnitudes.
\item Random Gaussian photometric errors are added to the magnitudes
  of each object based on the S/N given the limiting
  magnitudes of the simulated survey.
\item These noisy magnitudes are compared to the limiting magnitudes
  and all magnitudes that are fainter than the limits are considered
  non-detections.
\item Colour selection is performed with
  Eq.~(\ref{eq:colour_selection}) yielding the final source
  samples. Non-detections are used to set limits on the colour indices
  and also decide for those sources whether the selection criteria are
  satisfied or not.
\end{itemize}

These steps yield well-controlled photometric mock catalogues that can
be checked for number densities, mean magnitudes, and colour shifts in
the different scenarios mentioned above. These quantities can then be
compared to the theoretical predictions described in
Sect.~\ref{sec:theory}.

\section{Results}
\label{sec:results}
\subsection{Number-counts}
\begin{figure}
\includegraphics[width=\hsize]{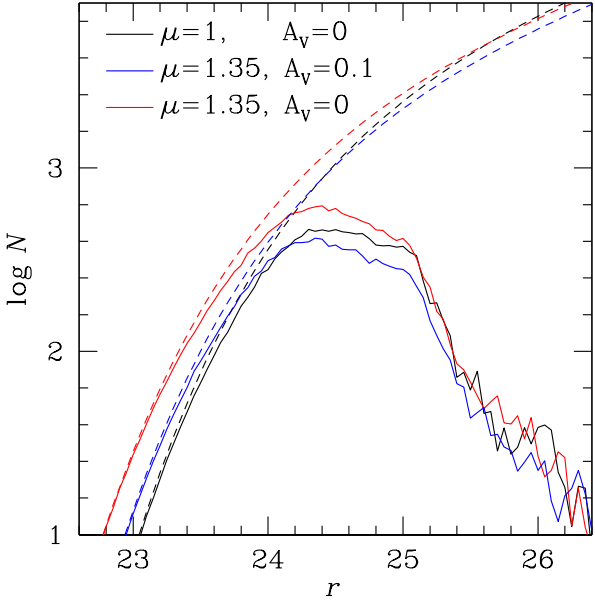}
\caption{Number-counts of the source LBG sample in the simulations
  (arbitrary normalisation) as a function of $r$-band magnitude. The
  dotted lines show the intrinsic number-counts whereas the solid
  lines include effects from photometric noise, source detection, and
  colour selection. Black lines represent an un-magnified sample, red
  lines a sample under a magnification of $\mu=1.35$, and blue lines a
  sample under a magnification of $\mu=1.35$ as well as dust
  absorption by a lens galaxy at redshift $z=0.7$ with an absorption
  in the rest-frame visual of $A_V=0.1$.}
\label{fig:nc}
\end{figure}

In Fig.~\ref{fig:nc} we show the input number-counts (dotted lines)
for the source sample as well as the number-counts including the
effects of noise, source detection, and colour selection. Three
different cases are displayed, the no-magnification and no-extinction
case, the magnification-only case, and the
magnification-plus-extinction case. For the setup here most of the
incompleteness is caused by the colour selection. This is due to the
fact that the depth of the $u$-, $g$-, and $r$-bands used for the
colour selection of the LBGs are limiting the depth of the LBG sample
more severely than the detection that is carried out in the deeper
$i$-band. This behaviour certainly depends on the relative depths of
the different bands and on the explicit form of the colour
selection. Some small amount of Eddington bias from photometric noise
and source detection is present in the magnitude range
$25<r<26$. However, this Eddington bias is unimportant in practice
since the depth of the $u$-, $g$-, and $r$-bands limits the useful
magnitude range after colour selection to $r\la25$.

One important aspect apparent from this plot is the fact that the
observational selection function is fixed in apparent
magnitude. Although the red lines (magnification-only scenario) are
shifted by more than 0.2mag horizontally\footnote{Note that the
  apparent horizontal shift is smaller than $2.5\log(1.35)\approx0.33$
  because of the additional suppression in vertical direction.} with
respect to the black lines (no-magnification and no-extinction) at the
bright end, the peak and the faint magnitude fall-off are almost
identical magnitudes. This might seem obvious but needs to be stressed
since it could potentially have an effect on how to interpret flux
magnification measurements in regimes where the source sample is
incomplete. In other words, the selection function is fixed in
apparent magnitudes.

\subsection{Magnification bias}
Equation~(\ref{eq:nc_mu_alpha}) relates the observed number-counts to
the intrinsic number-counts via the logarithmic slope of those
intrinsic number-counts, $\alpha_{\rm nc}(m)$
(Eq.~\ref{eq:alpha_nc}). Hence, the interpretation of magnification
bias measurements requires a precise knowledge of this slope. As will
be shown in the following there are situations where the real change
in number density under magnification can differ from the theoretical
expectations given in Eq.~(\ref{eq:nc_mu_alpha}). This happens if
effects of noise and colour selection play a role and the observed
number-counts are used for measuring $\alpha_{\rm nc}$ which is then
used in the predictions without correcting these effects.

\begin{figure}
\includegraphics[width=\hsize]{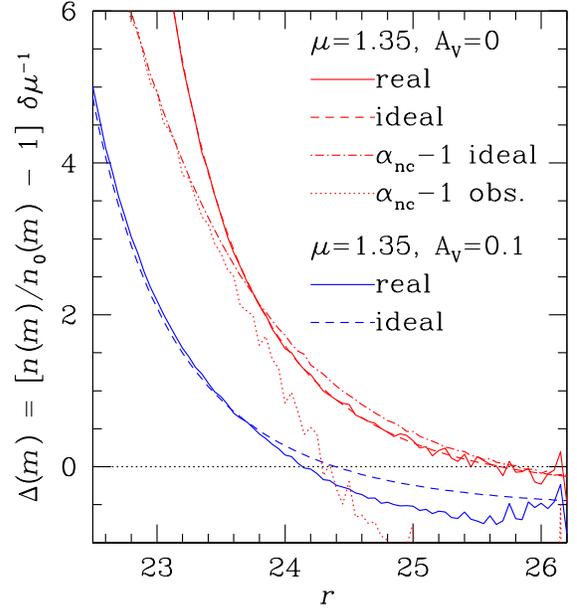}
\caption{Magnification strength as a function of $r$-band magnitude
  for a magnification of $\mu=1.35$. Shown is the ratio of the density
  change and the magnification excess $\delta\mu$. The dot-dashed and
  dotted red lines are predictions based on the weak lensing
  approximation estimated from the slope of the dotted and solid red
  curves in Fig.~\ref{fig:nc}, respectively, i.e. for perfect
  knowledge of the intrinsic number-counts and for the observed
  number-counts. The dashed red line represents the magnification
  strength of an ideal sample, i.e. without any systematic effects,
  whereas the solid red line represents the magnification strength
  when all systematic effects are considered. The blue dashed and
  solid lines show the same cases for a sample that is also affected
  by dust absorption by a lens galaxy at redshift $z=0.7$ with an
  absorption in the rest-frame visual of $A_V=0.1$.}
\label{fig:alpha}
\end{figure}

We define the magnification strength as
\begin{equation}
\Delta(m)=\left[\frac{n(m)}{n_0(m)}-1\right]\delta\mu^{-1}\,.
\end{equation}
This quantity equals $\alpha_{\rm nc}-1$ in the idealised case of
perfect knowledge of the number-counts and in the weak lensing regime.
Figure~\ref{fig:alpha} shows this magnification strength as a function
of apparent $r$-band magnitude for magnification only (here
$\mu=1.35$; red lines) and for another scenario that also includes
dust ($A_V=0.1$; blue lines). The red solid line represents the real
change in number density under magnification when all systematic
effects are included. The dashed red line represent the idealised case
without noise or incompleteness introduced by the object detection or
colour selection. This dashed red line is barely visible because it
agrees very well with the solid red line showing that the systematic
effects are rather unimportant in this particular case. In different
words, the incomplete sample represented by the solid red line behaves
identically (within noise) to the complete sample represented by the
dashed red line. This is a non-trivial result. It is not obvious from
Fig.~\ref{fig:nc} that the difference (because it is a logarithmic
plot) of the dashed black and red lines is the same as the difference
of the solid red and black lines in that figure. This finding also
means that one can get a good prediction of the magnification bias if
one has access to the intrinsic number-counts and one uses the slope
of those in Eq.~(\ref{eq:nc_mu_alpha}).

The dot-dashed red line is the weak lensing approximation, i.e. just
$\alpha_{\rm nc}-1$, using the slope of the intrinsic number-counts
(dashed black line in Fig.~\ref{fig:nc}). The difference between this
weak lensing approximation and the real magnification (solid red line
in Fig.~\ref{fig:alpha}) certainly increases for larger $\mu$. Here we
show a relatively large $\mu$ to illustrate this effect.

The dot-dashed red line in Fig.~\ref{fig:nc} assumes perfect knowledge
of the intrinsic number-counts. In contrast, using the observed
number-counts - uncorrected for any incompleteness - instead
(i.e. using the slope of the solid black line from Fig.~\ref{fig:nc})
to estimate $\alpha_{\rm nc}$ and predict the magnification strength
yields the dotted red line in Fig.~\ref{fig:nc} which deviates
strongly from the real situation at faint magnitudes. This means that
the observed number-counts are only usable in the magnitude regime
where the sample is complete and can not be used at the faint end
where incompleteness sets in. This incompleteness can come from source
detection, but in most practically relevant cases the incompleteness
from colour selection dominates.\footnote{This is actually a result of
  the typical survey design for multi-band imaging surveys. Usually
  one band is used as the detection band, and this band is
  considerably deeper than the other bands and/or taken under better
  observing conditions. The other bands used in the colour selection
  are shallower and hence introduce an incompleteness in a
  colour-selected galaxy sample (or in a photo-$z$ bin).}

The solid blue line shows the real $\Delta(m)$ in a scenario where the
lenses also contain some dust that absorbs part of the flux. It is
clear from this plot that even a moderate amount of dust in the lenses
has a profound effect on the observed density change and can not be
neglected when measuring magnification bias. Conversely, the
dependence of the magnification bias on apparent magnitude of the
background sample can be used to constrain the amount of dust in the
lenses and subsequently correct the lensing measurement for extinction
as was presented in \cite{2013MNRAS.429.3230H}. The dashed blue line
shows the density change in this scenario in the absence of noise and
incompleteness. Unlike in the magnification-only scenario where the
dashed and solid red lines lie exactly on top of each other, there is
a significant difference between the real and idealised cases for the
magnification plus dust-extinction scenario. The difference becomes
important for magnitudes fainter than the peak of the number-counts
(see Fig.~\ref{fig:nc}). This means that the magnification bias as a
function of magnitude is not just shifted to brighter magnitudes and
smaller values but it actually has a different dependence on magnitude
in the case with dust extinction than in the case without.

\subsection{Magnitude shift}
Next we explore how the theoretically calculated magnitude shift
(Eq.~\ref{eq:mag_shift_linear}) compares to the observed one in
different scenarios. We are concentrating on a relatively bright
$u$-dropout sample with $23<r<24.5$ here. This ensures that
essentially all objects are detected in all bands except for the
$u$-band, which makes the interpretation of the results much easier.

\begin{figure}
\includegraphics[width=\hsize]{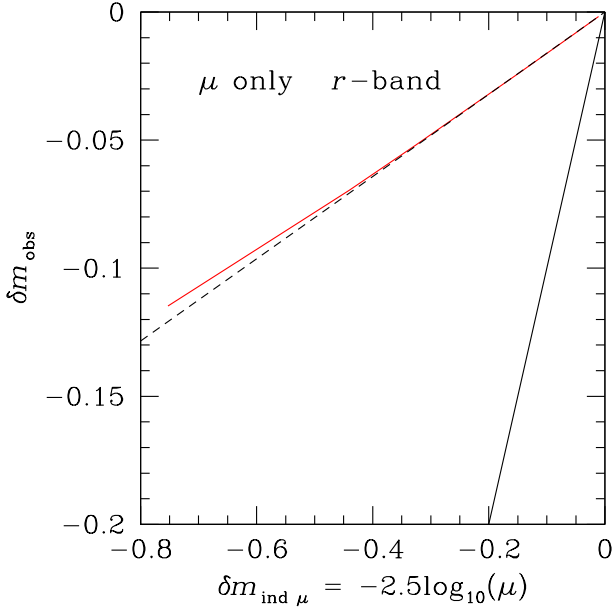}
\caption{Observed mean magnitude shift in the $r$-band as a function
  of the induced magnitude shift for a background sample with
  $23<r<24.5$ that is not affected by any systematic effects (red
  solid line). The weak lensing approximation is shown by the dashed
  line. The solid black line is the identity relation showing the
  strong suppression of the observed magnitude shift compared to the
  induced one experienced by an individual source (but
  unobservable). The constant $C_S$ corresponds to the slope of the
  dashed line.}
\label{fig:mag_shift_mu_only_ideal}
\end{figure}

Figure~\ref{fig:mag_shift_mu_only_ideal} shows the observed magnitude
shift as a function of the induced magnitude shift. Shown is the
idealised situation without noise or colour selection. In that case
the magnitude shift is achromatic so that we only show results for the
$r$-band in this figure. The solid red line is the actual magnitude
shift whereas the dashed black line represents the prediction from
Eqs.~\ref{eq:mag_shift_linear}~\&~\ref{eq:C_S} which assume the weak
lensing approximation. This approximation is working extremely well
for magnifications $\delta\mu\la50\%$, much better than for the
magnification bias!

\begin{figure}
\includegraphics[width=\hsize]{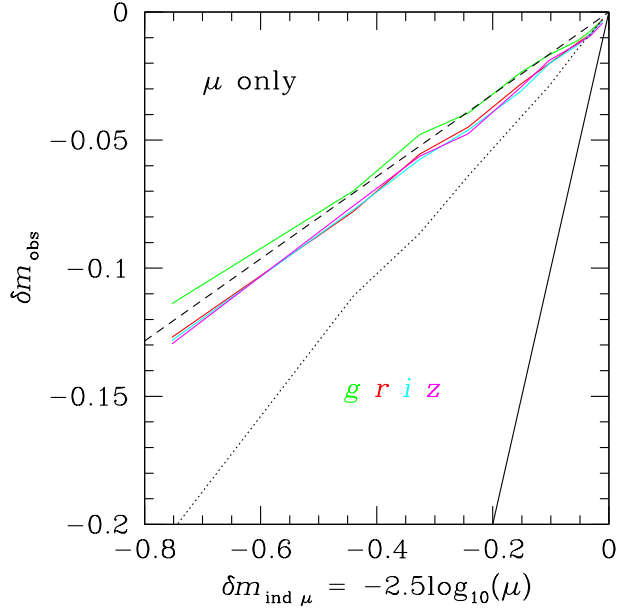}
\caption{Same as Fig.~\ref{fig:mag_shift_mu_only_ideal} but for a real
  sample that is affected by all the systematic effects described in
  Sect.~\ref{sec:systematics}. The differently coloured solid lines
  represent the results for the different bands. The systematic
  effects result in a chromaticity with the observed magnitude shift
  depending on wavelength. The dashed line is again the prediction
  from weak lensing theory under the assumption of perfect knowledge
  of the intrinsic number-counts, whereas the dotted line represents
  the same prediction but using the observed number-counts.}
\label{fig:mag_shift_mu_only_real}
\end{figure}

Figure~\ref{fig:mag_shift_mu_only_real} shows the magnitude shift
under more realistic conditions in four different bands (i.e. the
$griz$-bands, all of which are red-ward of the Lyman-break). Here
noise, object detection, colour selection, and a Gaussian distribution
in LBG colours are included. It is obvious that the observed magnitude
shift is not fully achromatic anymore.

In particular, the magnitude shift for the $g$-band differs from the
other bands by $\approx10\%$. This is due to the fact that the colour
selection in Eq.~(\ref{eq:colour_selection}) depends on the $g$-band
magnitude. The selection puts an upper limit on the $g-r$ colour of
$g-r\la1$, somewhat dependent on the $u-g$ colour.  In combination
with the $r$-band limit of $r<24.5$ also the $g$-band magnitude is
limited. This alters the $g$-band number-counts in a characteristic
way and suppresses some of the magnitude shift in that band. At the
same time it slightly increases the curvature of the $r$-band
number-counts.

For these reasons the theoretical prediction for the magnitude shift
(represented by the dashed black line) lies in between the observed
results for the $g$- and $r$-bands, with the $r$-band magnitude shift
slightly enhanced and the $g$-band magnitude shift slightly suppressed
with respect to the estimate. The prediction is based on the intrinsic
$r$-band number-counts. Using the observed $r$-band number-counts is
not a solution to correct for this bias since those are affected by
incompleteness (compare the dashed and solid black lines in
Fig.~\ref{fig:nc}). Hence the prediction with those observed
number-counts, represented by the dotted line, greatly over-predicts
the amplitude of the magnitude shift.

The case of the $g$-band is very different from the case of the $i$-
and $z$-bands, which are not used in selecting the sample. Their
number-counts are not limited in any way and they just follow the
behaviour of the $r$-band which is used to select the sample. This is
due to the fact that the $r-i$ and $r-z$ colour distributions do not
change under magnification whereas the $g-r$, $g-i$, and $g-z$ colour
distributions do change because of the selection effects described
above.

The strength of the suppression of the $g$-band magnitude shift and
the enhancement of the $r$-band magnitude shift compared to
theoretical expectations depends on the exact shape of the
number-counts. We tested power law distributions which theoretically
should not show any magnitude shift. This still holds even for a noisy
colour selected sample as the one used here. Only if there is some
curvature in the logarithmic number-counts in combination with a
colour selection can such a behaviour develop.

\begin{figure}
\includegraphics[width=\hsize]{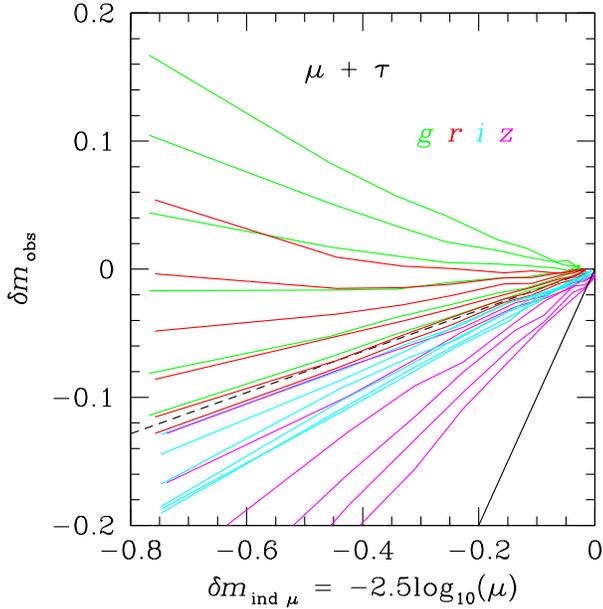}
\caption{Same as Fig.~\ref{fig:mag_shift_mu_only_real} but including
  varying amounts of dust in a lens galaxy at redshift $z=0.7$. It is
  obvious that the relationship between the induced and observed
  magnitude shift is extremely sensitive to the amount of dust
  absorption. Here we vary the scaling between the magnification
  excess and the absorption in the rest-frame visual,
  $A_V=c_{\rm d}\,\delta\mu$. The different lines correspond to
  different values of $c_{\rm d}=0.0;0.1;0.3;0.5;0.7;0.9$ going from
  bottom to top for the $g$- and $r$-bands and from top to bottom for
  the $i$- and $z$-bands, since objects are selected in $r$.}
\label{fig:mag_shift_mutau_real}
\end{figure}

So far we looked at just magnification. If there is additional dust
extinction/absorption the observed magnitude shift can not be
converted into an estimate for the magnification directly. This is
illustrated in Fig.~\ref{fig:mag_shift_mutau_real} where variable
amounts of extinction by Milky-Way like dust in a lens galaxy at
$z=0.7$ were added. The relation between observed and induced
magnitude shift is very sensitive to the value of $c_{\rm d}$ (see
Eq.~\ref{eq:A_V}) which differs for the different lines of each colour
in Fig.~\ref{fig:mag_shift_mutau_real}. Hence in a realistic scenario
the effect of dust has to be removed before magnification can be
measured with the magnitude shift.

One way of estimating and removing the effect of dust is to use colour
information. In the following section we explore how well this works
in the presence of selection effects.

\subsection{Colour excess}
\begin{figure}
\includegraphics[width=\hsize]{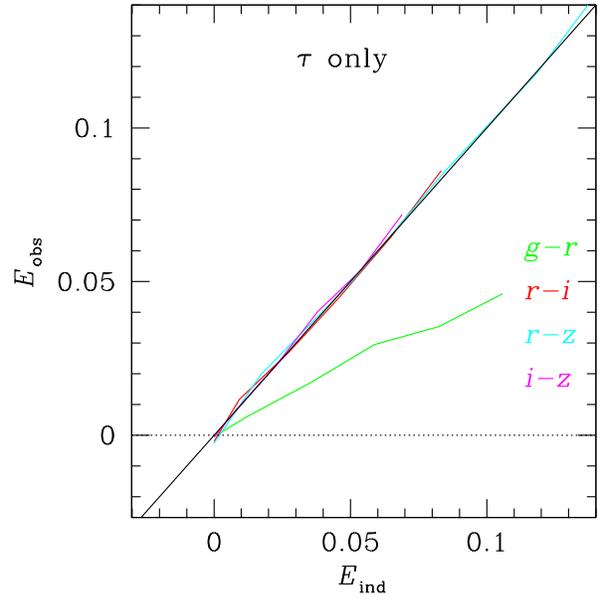}
\caption{Observed colour excess as a function of induced colour excess
  for a sample that is just affected by dust and no magnification. The
  differently coloured lines correspond to different colour indices,
  and the black line is the identity relation. Whereas the observed
  excess in the $r-i$, $r-z$, and $i-z$ colour indices can be directly
  converted to the induced colour excess, the $g-r$ colour index
  behaves differently because of the colour selection.}
\label{fig:col_shift_tau}
\end{figure}

Looking at the colour excess of background galaxies to estimate the
amount of dust in galaxy halos along the line of sight was suggested
by \cite{2010MNRAS.405.1025M}. The chromaticity of the magnitude shift
induced by magnification in the presence of noise and colour selection
as shown in Fig.~\ref{fig:mag_shift_mu_only_real} could mean that a
direct estimate of the amount of dust from the colour excess might
also be biased. As illustrated in Fig.~\ref{fig:col_shift_tau}, even
without magnification, i.e. just considering dust extinction, the
observed colour excess can be different from the induced colour
excess. In this plot these two quantities are compared for the
dust-only case. The observed colour excess in the $g-r$ colour is
lower than the induced colour excess by dust extinction by about a
factor of two. This is due to the systematic effects discussed above
which lead to galaxies with different SEDs entering the sample than
leaving the sample when extinction is present. The $g-r$ colour
distribution is the only one that is limited at $g-r\la1$ because of
the selection in Eq~(\ref{eq:colour_selection}). The change in this
distribution, which is the observable here, is slightly suppressed
because of this cut in the observed colour. Interestingly, the same
behaviour is observed even if we change the shape of the $r$-band
number-counts into a power law.

\begin{figure}
\includegraphics[width=\hsize]{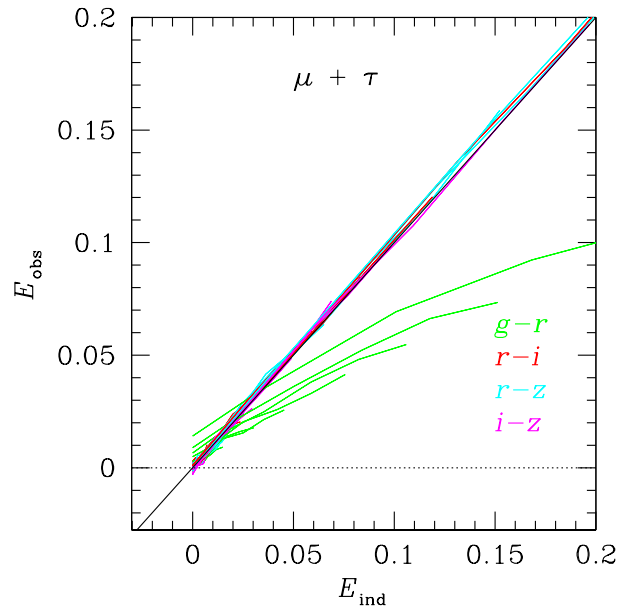}
\caption{Same as Fig.~\ref{fig:col_shift_tau} but additionally
  including varying amounts of magnification,
  $\mu=1.01; 1.03; 1.05; 1.10; 1.15; 1.25; 1.35; 1.5; 2.0$ going from
  bottom to top.  The observed excess in the $r-i$, $r-z$, and $i-z$
  colour indices remains a reliable estimate of the induced colour
  excess since the relation is independent of the amount of
  magnification. This means that the colour excess can be converted to
  an absorption which can be used to correct the magnitude shift for
  dust effects.}
\label{fig:col_shift_mutau}
\end{figure}

If on top of the dust extinction there is additional magnification the
situation changes only slightly. As can be seen in
Fig.~\ref{fig:col_shift_mutau} the observed colour excess in $r-i$,
$r-z$, and $i-z$ is still an unbiased estimate of the induced colour
excess. Due to the apparent chromaticity of the magnification-induced
magnitude shift in the $g$-band the observed colour excess in $g-r$
depends on the amount of magnification, with the different green lines
in Fig.~\ref{fig:col_shift_mutau} corresponding to different amounts
of magnification.

The result from Fig.~\ref{fig:col_shift_mutau} means that one can use
the $r-i$, $r-z$, and $i-z$ colour indices to directly (i.e. even in
the presence of selection effects and noise) estimate the colour
excess induced by dust. Assuming an extinction law \citep[or even
fitting the extinction law with multiple colour indices as it was done
in][]{2010MNRAS.405.1025M} one can convert the colour excess into an
absorption and correct the magnitude shift for dust extinction. Hence
one can recover the magnification-only results from
Fig.~\ref{fig:mag_shift_mu_only_real} by combining the measurements of
Fig.~\ref{fig:mag_shift_mutau_real} and
Fig.~\ref{fig:col_shift_mutau}.

\section{Discussion}
\label{sec:discussion}
The results discussed in this paper suggest that the interpretation of
flux magnification observables under realistic conditions can be more
complicated than previously thought. If effects like noise, colour
selection, and dust extinction are simulated the results can differ
greatly from theoretical predictions that ignore those effects. Here
we illustrate this on a source sample that is quite sensitive to those
systematic effects. In different situations these systematics could
possibly be more or less important. Dedicated simulations are
necessary to decide this and establish robust corrections for each
flux magnification measurement.

Given the importance of dust extinction in magnification bias as well
as magnitude shift measurements it is clear that any magnification
study has to take dust into account. A possible way to do this is to
use the colour excess from colour indices that are free from
systematics (like the three redder colour indices in our example) to
estimate the amount of extinction. A different and possibly
complementary way would be to use the dependence of the magnification
bias as a function of magnitude (Fig.~\ref{fig:alpha}). This technique
was applied in \cite{2013MNRAS.429.3230H} and is also generally useful
if colour information is unavailable or limited.

Once a robust estimate of the dust extinction is obtained one can
correct magnitude shift or magnification bias measurements for
this. Then one is still left with possible systematic errors from
incompleteness due to colour selection or source detection. In order
to account for these one has to revert to simulations. There does not
seem to be a purely observational way to establish these corrections
given the complexity of the data analysis involved (data reduction,
source detection, multi-colour photometry, photo-$z$, etc.).

\section{Summary and conclusions}
\label{sec:summary}
In this paper we simulate a high-redshift galaxy sample that has been
used in many magnification studies. By controlling several systematic
effects in these simulations we gain knowledge about their importance
in typical flux magnification science applications. We show that the
magnification bias for an incomplete sample is virtually identical to
the magnification bias of a complete sample. However, an incomplete
sample can not be used to estimate the slope of the magnitude
number-counts which is required for any prediction of flux
magnification. This estimate has to come from deeper data that are
unaffected by noise in the magnitude range of interest. We also show
that dust extinction can alter the magnification bias significantly
and needs to be taken into account when studying sources that are
magnified by lenses that potentially contain dust.

Magnitude shift measurements can become chromatic due to selection
effects, even when the underlying process - gravitational lensing - is
inherently achromatic. This chromaticity can mimic dust extinction and
needs to be distinguished from the latter by the techniques described
here. Similar to the magnification bias dust can affect the magnitude
shift and its effect needs to be modelled/removed before a lensing
measurement can be used to e.g. study dark matter halos.

The work presented here has potential implications for previous
studies using flux magnification. We simulated a CFHTLenS-like survey
with one particular source sample, $z\sim3$ $u$-dropouts, and can not
make strong general statements about results presented in the
literature. However, it is conceivable that some of the results need
to be revisited and checked for systematic effects. In the future it
will be indispensable to run such simulations if magnification is
going to be used as a precision tool in extragalactic astronomy
alongside weak lensing shear and other techniques.

The simulations used here are based on several simplifications. In
future studies these simplifications need to be dropped and the
simulations made more realistic. In particular, a full redshift and
SED range needs to be included and also a photo-$z$ code needs to be
run on the mock galaxies. This will then also allow galaxies at low
redshift to scatter into our high-redshift source sample, an effect
that was ignored here.

So far the simulations we are using are based purely on photometric
mock catalogues. For some precision studies and for analysing the
impact of some other systematic effects it will become necessary to
also include effects of galaxy clustering, blending, etc. Such
advanced mocks would have to be based on N-body simulations, possibly
with ray-tracing. Such a setup would come closer to a full end-to-end
treatment of magnification and this should be the long-term goal to
exploit this observational technique in very large surveys like
Euclid, LSST, and WFIRST.

\section*{Acknowledgements}
The author would like to thank Catherine Heymans, Peter Schneider,
Malte Tewes, Alan Heavens and Christopher Morrison for useful comments
on the paper draft.

H. Hildebrandt is supported by an Emmy Noether grant (No. Hi 1495/2-1)
of the Deutsche Forschungsgemeinschaft.

\bibliographystyle{mn2e_mod}

\bibliography{magnification_simulations}

\begin{thebibliography}{27}
\expandafter\ifx\csname natexlab\endcsname\relax\def\natexlab#1{#1}\fi

\bibitem[{{Bauer} {et~al}\mbox{.}(2014){Bauer}, {Gazta{\~n}aga},
  {Mart{\'{\i}}}, \& {Miquel}}]{2014MNRAS.440.3701B}
{Bauer} A.~H., {Gazta{\~n}aga} E., {Mart{\'{\i}}} P., {Miquel} R., 2014,
  \mnras, 440, 3701

\bibitem[{{Bertin} \& {Arnouts}(1996)}]{1996A&AS..117..393B}
{Bertin} E., {Arnouts} S., 1996, \aaps, 117, 393

\bibitem[{{Cardelli} {et~al}\mbox{.}(1989){Cardelli}, {Clayton}, \&
  {Mathis}}]{1989ApJ...345..245C}
{Cardelli} J.~A., {Clayton} G.~C., {Mathis} J.~S., 1989, \apj, 345, 245

\bibitem[{{Eddington}(1913)}]{1913MNRAS..73..359E}
{Eddington} A.~S., 1913, \mnras, 73, 359

\bibitem[{{Erben} {et~al}\mbox{.}(2013){Erben}, {Hildebrandt}, {Miller}, {van
  Waerbeke}, {Heymans}, {Hoekstra}, {Kitching}, {Mellier}, {Benjamin}, {Blake},
  {Bonnett}, {Cordes}, {Coupon}, {Fu}, {Gavazzi}, {Gillis}, {Grocutt}, {Gwyn},
  {Holhjem}, {Hudson}, {Kilbinger}, {Kuijken}, {Milkeraitis}, {Rowe},
  {Schrabback}, {Semboloni}, {Simon}, {Smit}, {Toader}, {Vafaei}, {van Uitert},
  \& {Velander}}]{2013MNRAS.433.2545E}
{Erben} T. {et~al.}, 2013, \mnras, 433, 2545

\bibitem[{{Ford} {et~al}\mbox{.}(2014){Ford}, {Hildebrandt}, {Van Waerbeke},
  {Erben}, {Laigle}, {Milkeraitis}, \& {Morrison}}]{2014MNRAS.439.3755F}
{Ford} J., {Hildebrandt} H., {Van Waerbeke} L., {Erben} T., {Laigle} C.,
  {Milkeraitis} M., {Morrison} C.~B., 2014, \mnras, 439, 3755

\bibitem[{{Ford} {et~al}\mbox{.}(2012){Ford}, {Hildebrandt}, {Van Waerbeke},
  {Leauthaud}, {Capak}, {Finoguenov}, {Tanaka}, {George}, \&
  {Rhodes}}]{2012ApJ...754..143F}
{Ford} J. {et~al.}, 2012, \apj, 754, 143

\bibitem[{{Giavalisco}(2002)}]{2002ARA&A..40..579G}
{Giavalisco} M., 2002, \araa, 40, 579

\bibitem[{{Heavens} \& {Joachimi}(2011)}]{2011MNRAS.415.1681H}
{Heavens} A.~F., {Joachimi} B., 2011, \mnras, 415, 1681

\bibitem[{{Heymans} {et~al}\mbox{.}(2012){Heymans}, {Van Waerbeke}, {Miller},
  {Erben}, {Hildebrandt}, {Hoekstra}, {Kitching}, {Mellier}, {Simon},
  {Bonnett}, {Coupon}, {Fu}, {Harnois D{\'e}raps}, {Hudson}, {Kilbinger},
  {Kuijken}, {Rowe}, {Schrabback}, {Semboloni}, {van Uitert}, {Vafaei}, \&
  {Velander}}]{2012MNRAS.427..146H}
{Heymans} C. {et~al.}, 2012, \mnras, 427, 146

\bibitem[{{Hildebrandt} {et~al}\mbox{.}(2012){Hildebrandt}, {Erben}, {Kuijken},
  {van Waerbeke}, {Heymans}, {Coupon}, {Benjamin}, {Bonnett}, {Fu}, {Hoekstra},
  {Kitching}, {Mellier}, {Miller}, {Velander}, {Hudson}, {Rowe}, {Schrabback},
  {Semboloni}, \& {Ben{\'{\i}}tez}}]{2012MNRAS.421.2355H}
{Hildebrandt} H. {et~al.}, 2012, \mnras, 421, 2355

\bibitem[{{Hildebrandt} {et~al}\mbox{.}(2011){Hildebrandt}, {Muzzin}, {Erben},
  {Hoekstra}, {Kuijken}, {Surace}, {van Waerbeke}, {Wilson}, \&
  {Yee}}]{2011ApJ...733L..30H}
{Hildebrandt} H. {et~al.}, 2011, \apjl, 733, L30+

\bibitem[{{Hildebrandt} {et~al}\mbox{.}(2009{\natexlab{a}}){Hildebrandt},
  {Pielorz}, {Erben}, {van Waerbeke}, {Simon}, \&
  {Capak}}]{2009A&A...498..725H}
{Hildebrandt} H., {Pielorz} J., {Erben} T., {van Waerbeke} L., {Simon} P.,
  {Capak} P., 2009{\natexlab{a}}, \aap, 498, 725

\bibitem[{{Hildebrandt} {et~al}\mbox{.}(2009{\natexlab{b}}){Hildebrandt}, {van
  Waerbeke}, \& {Erben}}]{2009A&A...507..683H}
{Hildebrandt} H., {van Waerbeke} L., {Erben} T., 2009{\natexlab{b}}, \aap, 507,
  683

\bibitem[{{Hildebrandt} {et~al}\mbox{.}(2013){Hildebrandt}, {van Waerbeke},
  {Scott}, {B{\'e}thermin}, {Bock}, {Clements}, {Conley}, {Cooray}, {Dunlop},
  {Eales}, {Erben}, {Farrah}, {Franceschini}, {Glenn}, {Halpern}, {Heinis},
  {Ivison}, {Marsden}, {Oliver}, {Page}, {P{\'e}rez-Fournon}, {Smith},
  {Rowan-Robinson}, {Valtchanov}, {van der Burg}, {Vieira}, {Viero}, \&
  {Wang}}]{2013MNRAS.429.3230H}
{Hildebrandt} H. {et~al.}, 2013, \mnras, 429, 3230

\bibitem[{{M{\'e}nard} {et~al}\mbox{.}(2010){M{\'e}nard}, {Scranton},
  {Fukugita}, \& {Richards}}]{2010MNRAS.405.1025M}
{M{\'e}nard} B., {Scranton} R., {Fukugita} M., {Richards} G., 2010, \mnras,
  405, 1025

\bibitem[{{Miller} {et~al}\mbox{.}(2013){Miller}, {Heymans}, {Kitching}, {van
  Waerbeke}, {Erben}, {Hildebrandt}, {Hoekstra}, {Mellier}, {Rowe}, {Coupon},
  {Dietrich}, {Fu}, {Harnois-D{\'e}raps}, {Hudson}, {Kilbinger}, {Kuijken},
  {Schrabback}, {Semboloni}, {Vafaei}, \& {Velander}}]{2013MNRAS.429.2858M}
{Miller} L. {et~al.}, 2013, \mnras, 429, 2858

\bibitem[{{Morrison} {et~al}\mbox{.}(2012){Morrison}, {Scranton}, {M{\'e}nard},
  {Schmidt}, {Tyson}, {Ryan}, {Choi}, \& {Wittman}}]{2012MNRAS.426.2489M}
{Morrison} C.~B., {Scranton} R., {M{\'e}nard} B., {Schmidt} S.~J., {Tyson}
  J.~A., {Ryan} R., {Choi} A., {Wittman} D.~M., 2012, \mnras, 426, 2489

\bibitem[{{Scranton} {et~al}\mbox{.}(2005){Scranton}, {M{\'e}nard}, {Richards},
  {Nichol}, {Myers}, {Jain}, {Gray}, {Bartelmann}, {Brunner}, {Connolly},
  {Gunn}, {Sheth}, {Bahcall}, {Brinkman}, {Loveday}, {Schneider}, {Thakar}, \&
  {York}}]{2005ApJ...633..589S}
{Scranton} R. {et~al.}, 2005, \apj, 633, 589

\bibitem[{{Steidel} {et~al}\mbox{.}(2003){Steidel}, {Adelberger}, {Shapley},
  {Pettini}, {Dickinson}, \& {Giavalisco}}]{2003ApJ...592..728S}
{Steidel} C.~C., {Adelberger} K.~L., {Shapley} A.~E., {Pettini} M., {Dickinson}
  M., {Giavalisco} M., 2003, \apj, 592, 728

\bibitem[{{Steidel} {et~al}\mbox{.}(1996){Steidel}, {Giavalisco}, {Pettini},
  {Dickinson}, \& {Adelberger}}]{1996ApJ...462L..17S}
{Steidel} C.~C., {Giavalisco} M., {Pettini} M., {Dickinson} M., {Adelberger}
  K.~L., 1996, \apjl, 462, L17

\bibitem[{{Teerikorpi}(2004)}]{2004A&A...424...73T}
{Teerikorpi} P., 2004, \aap, 424, 73

\bibitem[{{Umetsu} {et~al}\mbox{.}(2011){Umetsu}, {Broadhurst}, {Zitrin},
  {Medezinski}, \& {Hsu}}]{2011ApJ...729..127U}
{Umetsu} K., {Broadhurst} T., {Zitrin} A., {Medezinski} E., {Hsu} L.-Y., 2011,
  \apj, 729, 127

\bibitem[{{Umetsu} {et~al}\mbox{.}(2014){Umetsu}, {Medezinski}, {Nonino},
  {Merten}, {Postman}, {Meneghetti}, {Donahue}, {Czakon}, {Molino}, {Seitz},
  {Gruen}, {Lemze}, {Balestra}, {Ben{\'{\i}}tez}, {Biviano}, {Broadhurst},
  {Ford}, {Grillo}, {Koekemoer}, {Melchior}, {Mercurio}, {Moustakas}, {Rosati},
  \& {Zitrin}}]{2014ApJ...795..163U}
{Umetsu} K. {et~al.}, 2014, \apj, 795, 163

\bibitem[{{van der Burg} {et~al}\mbox{.}(2010){van der Burg}, {Hildebrandt}, \&
  {Erben}}]{2010A&A...523A..74V}
{van der Burg} R.~F.~J., {Hildebrandt} H., {Erben} T., 2010, \aap, 523, A74

\bibitem[{{van Waerbeke}(2010)}]{2010MNRAS.401.2093V}
{van Waerbeke} L., 2010, \mnras, 401, 2093

\bibitem[{{Van Waerbeke} {et~al}\mbox{.}(2010){Van Waerbeke}, {Hildebrandt},
  {Ford}, \& {Milkeraitis}}]{2010ApJ...723L..13V}
{Van Waerbeke} L., {Hildebrandt} H., {Ford} J., {Milkeraitis} M., 2010, \apjl,
  723, L13

\end{thebibliography}

\label{lastpage}
\end{document}